\documentclass[aps,prd,showpacs,preprintnumbers,nofootinbib,twocolumn]{revtex4}

\newcommand{\ga}{\alpha} 
\newcommand{\gb}{\beta} 
\renewcommand{\gg}{\gamma} 
\newcommand{\gd}{\delta} 
\renewcommand{\ge}{\epsilon}

\newcommand{\gvf}{\varphi} 
\newcommand{\gc}{\chi} 
\newcommand{\gx}{\xi} 
\newcommand{\gm}{\mu}  
\newcommand{\gn}{\nu} 
 
\newcommand{\gk}{\kappa} 
\newcommand{\gl}{\lambda} 
\newcommand{\gr}{\rho} 
\newcommand{\gth}{\theta} 
\newcommand{\gs}{\sigma} 
 
\newcommand{\go}{\omega}

\newcommand{\gps}{\psi}

\newcommand{\gG}{\Gamma} 
\newcommand{\gD}{\Delta} 
\newcommand{\gF}{\Phi}

\newcommand{\gL}{\Lambda} 
\newcommand{\gS}{\Sigma} 
 
\newcommand{\gO}{\Omega} 
 
\newcommand{\gPs}{\Psi} 
 
\newcommand{\cK}{{\cal K}} 
\newcommand{\cL}{{\cal L}} 
\newcommand{\cM}{{\cal M}}

\newcommand{\uI}{{\underline I}}

\newcommand{\tk}{{\tilde k}}

\newcommand{\tA}{{\tilde A}}

\newcommand{\uga}{{\underline\alpha}} 
\newcommand{\ugb}{{\underline\beta}}

\newcommand{\bh}{{\bar h}}

\newcommand{\bk}{{\bar k}}

\newcommand{\bz}{{\bar z}} 
\newcommand{\bA}{{\bar A}}

\newcommand{\bF}{{\bar F}}

\newcommand{\bM}{{\bar M}} 
\newcommand{\bN}{{\bar N}}

\newcommand{\bW}{{\bar W}}

\newcommand{\bge}{{\bar\epsilon}}

\newcommand{\bgvf}{{\bar\varphi}} 
\newcommand{\bgc}{{\bar\chi}}

\newcommand{\bgl}{{\bar\lambda}}

\newcommand{\bgo}{{\bar\omega}}

\newcommand{\bgps}{{\bar\psi}}

\newcommand{\bgF}{{\bar\Phi}}

\newcommand{\bgL}{{\bar\Lambda}} 
\newcommand{\bgS}{{\bar\Sigma}}

\newcommand{\bgPs}{{\bar\Psi}} 
 
\newcommand{\bcM}{{\bar{\cal M}}}

\newcommand{\tr}{\text{tr}} 
 
\newcommand{\Id}{\text{\small 1}\hspace{-3.5pt}\text{1}}
\newcommand{\slashed}{\hspace{-1.1ex}/}

\newcommand{\der}{\partial}

\newcommand{\sder}{\der\slashed}

\newcommand{\nit}{\noindent}

\newcommand{\ct}{\cite} 
\newcommand{\bit}{\bibitem}

\newcommand{\hs}[1]{\hspace{#1 em}}
\newcommand{\labl}[1]{\label{#1}} 
\newcommand{\Kh}{K\"{a}hler}  
\newcommand{\beq}{\begin{equation}} 
\newcommand{\feq}{\end{equation}} 
\newcommand{\barr}{\begin{array}} 
\newcommand{\earr}{\end{array}}

\begin{document}

\title{Gauged supersymmetric $\gs$--models and soft breaking terms}
\author{T.S.\ Nyawelo}
\email[]{tnyawelo@ictp.trieste.it.} 

\affiliation{The Abdus Salam ICTP, Strada Costiera 11, I-34014 Trieste, Italy.}


\begin{abstract}
Supersymmetric non--linear $\gs$--models in four dimensions with $D$--term potentials can sometimes have a singular metric. As the kinetic terms of scalar fields and their chiral fermionic 
partners are determined by this metric, it follows that their kinetic energy vanishes in the vacuum.
In previous work we have shown for a simple model that this degeneracy of the sigma-model metric can be lifted by soft supersymmetry breaking terms. In this letter I introduce soft breaking terms in  more realistic models based on $SO(10)$ and $E_6$ and compute their resulting mass spectra.
\end{abstract}
\maketitle

\vskip1cm

The kinetic part of a lagrangian for chiral multiplets in supersymmetric 
sigma-models in four dimensions is determined by the metric of the \Kh\ manifold \ct{zumino,freed-alg}.  However, it is not automatically guaranteed that this metric is always invertible. 
In previous work  \cite{STJ}, it was found that the \Kh\ metric develops zero-modes in the vacuum: the metric for the Goldstone bosons and their fermionic partners vanishes. In \cite{TsnFrJvSgn} we have shown for a simple model that soft supersymmetry breaking terms can shift the minimum away from the singular point and provide a consistent mass-spectrum. In this letter, I introduce soft supersymmetry breaking terms in more 
realistic models based on the coset spaces $E_6/[SO(10)\times U(1)]$ and 
$SO(10)/U(5)$.

From the point of view of unification the coset space 
$SO(10)/[SU(5)\times U(1)]$ is a very interesting for  phenomenological 
applications as both $SO(10)$ 
and $SU(5)$ are often used GUT groups \ct{Ong}. Since supersymmetric
pure $\gs$-models on coset spaces $G/H$ of K\"{a}hler type, are known to be anomalous 
\ct{moor-nel,coh-gom,buch-ler,KotchShore}, a supersymmetric model built on 
the  
$SO(10)/[SU(5)\times U(1)]$ is 
not free of anomalies by itself as all the $\underline{10}$ anti-symmetric 
complex coordinates $z^{ij}$ and their chiral superpartners $\gps^{ij}_L$ ($i, j = 1,\dots, 5)$ of this manifold carry the same  
charges. To construct a consistent supersymmetric model on this coset one has to 
include the fermionic partners of the  coordinates in an anomaly-free 
representation.  As $SU(5)$ representations are not anomaly free by themselves, 
we have to use the full $SO(10)$ representations for our additional matter 
coupling in this case. This has been achieved in \cite{STJ} by introducing 
a singlet $\underline{1}$ and a completely anti-symmetric tensor with 4 
indices, 
which is equivalent to $\underline{\bar{5}}$ to complete the set of complex 
chiral 
superfields to form a $\underline{16}$ of $SO(10)$. The anti-symmetric 
coordinates of the coset are combined into a $\underline{10}$ of $SU(5)$ 
with a unit $U(1)$ charge. An anomaly free representation is obtained using the 
branching of the $\underline{16}$. Indeed, its decomposition under $SU(5)$ 
reads: $\underline{16} = \underline{10}(1) + \underline{\bar{5}}(-3) + 
\underline{1}(5)$, where the numbers in parentheses denote the relative $U(1)$ charges. Therefore, the supersymmetric model on $SO(10)/U(5)$ is defined 
by three chiral superfields 
$(\gF^{ij},\gPs_i,\gPs)$: the target manifold $SO(10)/U(5)$ is parametrized by 10 
anti-symmetric complex fields $z^{ij}$ in a chiral 
superfield $\gF^{ij} = (z^{ij}, \gps^{ij}_L, H^{ij})$, to which are added $SU(5)$ 
vector and scalar matter multiplets denoted respectively 
by $\gPs_i = (k_i,\go_{L~i}, B_i)$, and $\gPs = (h,\varphi_L, F)$. 

The complete \Kh\ potential of the model is obtained using the method developed in \ct{BKMU,BKMU1}: 
\begin{eqnarray}
\cK(z, \bz; k, \bk; h, \bh) &=&  \frac{1}{2f^2}\ln\det\gc^{-1} + 
(\det\gc)^2|h|^2\nonumber\\
[2mm]
& & + (\det\gc)^{-1}k\gc^{-1}\bk
\labl{comkh}
\end{eqnarray}
with the submetric $\gc^{-1} = \Id_5 + f^2 z \bz$. The dimensionful  constant $f$ is introduced to assign correct physical dimensions 
to the scalar fields $(z,\bz)$. The \Kh\ metric derived from this 
\Kh\ potential $\cK$ possesses a set of holomorphic Killing vectors generating a 
non-linear representation of $SO(10)$:
\begin{eqnarray}
\gd z &=& \frac 1f\,  x - u^Tz - zu + f\,  zx^\dag z,\nonumber\\
[2mm]
\gd  h &=& 2\tr(f\, zx^\dag - u^T)h,\nonumber\\
[2mm] 
\gd k &=&  - k \Bigl(-u^T + f\, zx^\dag  + 
\tr (-u^T + f\, zx^\dag) \Id_5\Bigl).
\labl{vkilling}
\end{eqnarray}
Here $u$ represents the parameters of the linear diagonal $U(5)$ 
transformations, 
and $(x, x^\dag )$ are 
the complex parameters of the broken off-diagonal $SO(10)$ 
transformations. 

In order for the chiral fermions $(\gps^{ij}_L,\go_{Li},\gvf_L)$ to have a 
physical interpretation as describing a family quarks and 
leptons, gauge interactions should be introduced. In this case supersymmetry 
implies the 
addition of a potential from  elimination of the auxiliary $D^i$ fields by
substitution for the Killing potentials \cite{bw,WessBagger}. We will consider 
two different options:
gauging of the full $SO(10)$ symmetry, and gauging of the stability 
subgroup $SU(5)\times U(1)$ only. We denote collectively the $SO(10)$ gauge fields as $A_\gm = (U_\gm,W^{\dag}_\gm, W_\gm)$  with $W^{\dag}_\gm$ and $W_\gm$ the gauge fields
corresponding to the broken $SO(10)$ transformations parametrized by 
$(x, x^\dag)$ and with $U_{\mu}$, the gauge field of the diagonal 
transformations parametrized by $u$.  This requires the introduction of covariant 
derivatives for the dynamical fields. We give here the covariant 
derivatives for the scalar fields only: 
\begin{eqnarray}
D_\gm z &=& \der_\gm z - g_{10}\Bigl(\frac{1}{f} W_\gm - U^T_\gm z - 
z\,U_\gm  
+ f z W_\gm{}^\dag z\Bigl),\nonumber\\
[2mm] 
D_\gm k &=& \der_\gm k + g_{10}\,k\,\Bigl(f W_\gm{}^\dag z - U^T_\gm\nonumber\\
[2mm] 
& &+ \tr(f W_\gm{}^\dag z - U^T_\gm )\Id_5\Bigl)
,\nonumber\\
D_\gm h &=& \der_\gm h - 2 g_{10}\tr\Bigl(f W_\gm{}^\dag z - U^T_\gm\Bigl)h.
\labl{CovDer} 
\end{eqnarray}

For the $D$--term scalar potential we need the $SO(10)$ Killing
potentials. The full Killing potential $\cM$ generating the Killing 
vectors  
(\ref{vkilling}) can be written as
\begin{eqnarray}
\cM(u,x^\dag, x) = \tr\Bigl(u\cM_u + x^\dag\cM_x + x\cM_x^\dag\Bigl),
\end{eqnarray}
%
%
where $\cM_u$ is the $U(5)$ Killing potentials, and ($\cM_x,\cM_{x^\dag}$) are the broken Killing potentials. The $U(1)$ Killing potential $\cM_Y$ is defined as the trace of 
$U(5)$ Killing potential $\cM_u$ whereas the remaining $SU(5)$ 
Killing potential $\cM_t$ is defined as a traceless part of $\cM_u$: 
\begin{eqnarray}
\cM_t = \cM_u - \frac 15 \cM_Y \, \Id_5, 
 \qquad \cM_Y = \tr \cM_u.
\end{eqnarray}

The coupling of the gauge multiplets to the supersymmetric non-linear $\gs$--model on 
$SO(10)/U(5)$ has interesting consequences for the spectrum. It can induce 
spontaneous breaking of  supersymmetry, and further spontaneous breaking of the 
internal symmetry. For example, if we gauge the full $SO(10)$, all the Goldstone 
bosons $(z,\bz)$ are absorbed by the vector bosons $(W^{\dag}_\gm, W_\gm)$ which 
become massive. In this case we may choose to study the model in the unitary 
gauge $ z = \bz = 0$. However, it was found in \cite{STJ} that in this gauge, 
the \Kh\ metric develops zero-modes in the vacuum: the metric ${G_\gs}_{z^{ij}\bz_{kl}}$ for the Goldstone bosons and their fermions vanishes. 

To see this, we start from the scalar potential.  As already 
stated, we choose the unitary gauge: $z = \bz = 0$. Then the potential\footnote{With the given field content, it is not possible to construct an
$SO(10)$-invariant trilinear superpotential. Thus the full potential contains only a $D$-term potential.} for the fully gauged $SO(10)$  model reads
\begin{eqnarray}
V_{\mathrm{unitary}} &=& \frac{g_{10}^2}{10}\Bigl(10|h|^2 - 
\frac{5}{2f^2} - 
6|k|^2\Bigl)^2 + 
\frac{2}{5}g_{10}^2\Bigl(|k|^2\Bigl)^2.
\labl{uni}
\end{eqnarray}
From this we see that we only have a supersymmetric minimum if 
\begin{eqnarray}
|k|^2 = 0,\quad |h|^2 = \frac{1}{4f^2}.
\end{eqnarray}
It can be seen immediately that this solution yields the vanishing of the 
\Kh\ metric:
\begin{eqnarray}
\langle{G_\gs}_{(ij)}{}^{(kl)} \rangle &=& \gd_{i}^{[k}\gd_{j}^{l]}\Bigl(\frac{1}{2f^2} - 
2|h|^2
+ 
|k_i|^2\Bigl)\nonumber\\
[2mm]
& & + k^{(k}\gd_{(i}{}^{l)}\bk_{j)} = 0.
\labl{vanishing}
\end{eqnarray}
In this case the kinetic terms of the Goldstone superfield components vanish, 
therefore, mass terms for the $SO(10)$ gauge fields  $(W^{\dag}_\gm, W_\gm)$ 
vanish as well. Moreover, the theory becomes strongly 
coupled, with some of the four--fermion interactions exploding, namely:  
\begin{eqnarray}
\cL_{\mathrm{4-ferm}} = R_{z \bz h \bh} \,  \bgps_R \gps_L \,  \bgvf_L \gvf_R 
+ {\rm perm.}
\labl{4fermi} 
\end{eqnarray}
with the curvature components given by  
\begin{eqnarray}
\langle R_{z \bz h \bh} \rangle &=&   \langle{R}_{(ij)}{}^{(kl)}{}_{h\bh} \rangle\\
[2mm]
&= &-2 f^2\gd_{i}^{[k}\gd_{j}^{l]}
\Bigl(1 +  2|h|^2(\frac{1}{2f^2} - 2|h|^2 )^{-1}\Bigl)\nonumber.
\label{4fermicouplings}
\end{eqnarray}
%
This may point to a restauration of the $SO(10)$ symmetry. However, not all of the physics behind this model is as yet
understood.

To avoid the problem of vanishing of the 
\Kh\ metric, we shift the minimum of the 
potential away from the singular point by adding $SO(10)$-invariant soft 
supersymmetry breaking scalars mass terms
\begin{eqnarray}
\gD V = \mu_1^2\,(\det\gc)^2|h|^2 + \mu_2^2\, (\det\gc)^{-1}k\gc^{-1}\bk
\labl{so10soft}
\end{eqnarray}
to the potential. As a result the minimum of the potential is shifted to a 
position where the 
expectation value of the K\"{a}hler metric is not
vanishing; and the scalar $h$ gets a 
vacuum expectation value 
\begin{eqnarray}
|k|^2 = 0,\quad |h|^2 = v^2 = \frac{1}{4f^2} - \frac{\gm_1^2}{20g_{10}^2},\quad \gm_1^2 < \frac{5g^2_{10}}{f^2},
\labl{MINIMUM}
\end{eqnarray}
breaking the linear local $U(1)$ subgroup. The corresponding $U(1)$ vector becomes massive; 
and the remaining vectors of $SU(5)$ stay 
massless. In the fermionic sector, two Dirac fermions are  realized as a 
combination of the fermions of the chiral multiplets with the gauginos.

We now present details of the above mass spectrum. Since in general $SO(10)$ is 
broken in the vacuum, the Goldstone bosons $(\bz, z)$ are absorbed in
the longitudinal component of the charged vector bosons, and 
we may choose the unitary gauge $\bz = z = 0$. Before computing the mass 
spectrum of the theory, we first decompose the $U(5)$ vector multiplet $U^i_{~j} = (U^i_{\gm j},\gL^i_{Rj})$ 
into a $U(1)$  and $SU(5)$ vector multiplets denoted respectively by 
$A = (A_\gm, \lambda_R)$  
and $V^i_{~j} = (V^i_{\gm j},\gl^i_{Rj})$:
\begin{eqnarray}
V = U - \frac{1}{5}A\Id_5\quad \tr (V) = 0,\quad A = \tr (U).
\labl{U5}
\end{eqnarray}
It follows that the kinetic terms for the $U(1)$ multiplet are not canonically normalized. To obtain the standard normalization,  we redefine the $U(1)$ multiplet according to
\begin{eqnarray}
A\rightarrow\sqrt{5}(\tilde{A_\gm}, \tilde{\lambda}_R).
\end{eqnarray}
With the redefined fields, the part of the lagrangian that determine the mass 
spectrum of the theory; in the unitary gauge $z = \bz = 0$ reads:
\begin{eqnarray}
\cL &=& - \gd_{i}^{[k}\gd_{j}^{l]}\,\frac{\gm^2_1}{10g^2_{10}}\Bigl(g^2_{10}W^{\dagger{(ij)}}\cdot W_{(kl)} + 
\bgps^{(ij)}_L\stackrel{\leftrightarrow}{\sder}\gps_{(kl)L}\Bigl)\nonumber\\
[2mm]
& &
- \Bigl(\frac{1}{2}(\der\gr)^2 + 20g^2_{10}v^2\gr^2
 + 20g^2_{10}v^2\tA^2 + 
 \bar{\varphi}_L \hs{-.2} \stackrel{\leftrightarrow}{\sder} \hs{-.2} \varphi_L\Bigl)
\nonumber\\
[2mm]
& &
 - \Bigl(\frac{1}{2}\der\tk\cdot\der\tk + \frac{1}{2f^2}(\frac{3\gm_1^2}{5} + \gm_2^2)\tk^2 + \bgo^i_L \hs{-.2} \stackrel{\leftrightarrow}{\sder} \hs{-.2} \go_{iL}\Bigl)
\nonumber\\
[2mm]
& & 
- \frac{1}{4}\Bigl[\frac{1}{2}\bF_{(ij)}(W)\cdot F^{(ij)}(W) +
F_{\gm\gn}^2(\tilde{A})\nonumber\\
[2mm]
& & + F^i\,_{j}(V)\cdot F^i\,_{j}(V)\Bigl]
- \tilde{\bgl}_R\stackrel{\leftrightarrow}{\sder}\tilde{\gl}_{R} - 
\bgl_{R}^i\,_{j}\stackrel{\leftrightarrow}{\sder}\gl_{R}^i\,_{j}\nonumber\\
[2mm]
& & - \frac{1}{2}\Bigl(\frac{1}{2}\bgl^{(ij)}_R\stackrel{\leftrightarrow}{\sder}\gl_{(ij)R} + \frac{1}{2}\bgl^{(ij)}_L\stackrel{\leftrightarrow}{\sder}\gl_{(ij)L}\Bigl)
\nonumber\\
[2mm]
& & 
+ 2\sqrt{2}g_{10}\,{G_\gs}_{(ij)}{}^{(kl)}\,\Bigl[\frac{1}{f}\bgl^{(ij)}_R\gps_{L(kl)} + 
\textrm{h.c.}\Bigl]\nonumber\\
[2mm]
& &  + 2\sqrt{2}g_{10}\Bigl[2\sqrt{5}v\tilde{\bgl}_R\gvf_L + \textrm{h.c.}\Bigl].
\labl{fulllag}
\end{eqnarray}
%
%
%
%
In this expression we have expand the full potential (\ref{uni}) 
and (\ref{so10soft})
 to second order in fluctuations $\gr$ and $\tk$ with scalar $\gr$ defined by $h =(v + \frac{1}{\sqrt{2}}\gr)e^{\frac{1}{\sqrt{2}v}i\ga}$ around the absolute minimum (\ref{MINIMUM}). The bosonic mass spectrum can be read off 
easily form the lagrangian (\ref{fulllag}); they read:
\begin{eqnarray}
{\mathrm{m}}^2_W &=& \frac{4}{f^2}\frac{\gm^2_1}{10},\quad{\mathrm{m}}^2_\gr = {\mathrm{m}}^2_\tA = 
40 g^2_{10}v^2,\nonumber\\
{\mathrm{m}}^2_\tk &=& 
\frac{1}{f^2}(\frac{3\gm_1^2}{5} + \gm_2^2).
\end{eqnarray}
In the fermionic sector, two Dirac fermions are  formed by combining the 
quasi-Goldstone fermions $\gps^{[ij]}_L$ and $\gvf_L$ 
with the right-handed gauginos $\gl^{[ij]}_{R}$ and $\tilde{\gl}_R$ 
\begin{eqnarray}
\gPs = \tilde{\gl}_R  + \gvf_L,\qquad\gL^{[ij]} = 
\frac{\gm_1}{g_{10}\sqrt{10}}\gps^{[ij]}_L + \frac{1}{2}\gl^{[ij]}_R,
\labl{diracspin}
\end{eqnarray}
with masses: $\mathrm{m}_{\gPs} = 2g_{10}v\sqrt{10}$ and $\mathrm{m}_{\gL} = \frac{\sqrt{2}\gm_1}{\sqrt{5}f}$. The $\underline{\bar{5}}$  of the left-handed chiral fermions 
$\go_{iL}$, the $\underline{10}$ of the left-handed gaugino's $\gl^{[ij]}_L$, and  the Majorana fermions $\gl^i_{Rj}$ that 
are the gauginos of the unbroken $SU(5)$ symmetry remain massless. Notice here that in the limit $\mu^2_{1,2}\rightarrow 0$ and $g_{10} = g_1$, one gets 
the same massive multiplets in the model with only gauged linear subgroup $SU(5)\times U(1)$. The only difference is that in the case of gauged linear subgroup $SU(5)\times U(1)$ there are 20 
massless Goldstone bosons $(\tilde{\bz},\tilde{z})$, and their superpartners 
$(\gps_L,\bgps_L)$; and no gauge bosons $(\bW, W)$ 
of the 20 broken generators of $SO(10)$. (We have observed a similar thing to 
happen also in $E_6/SO(10)\times U(1)$ model discussed below.) 

As an alternative to gauging $SO(10)$, one can gauge  only the linear subgroup 
$SU(5) \times U(1)$ instead. It is then allowed in this case to introduce a Fayet-Iliopoulos term with parameter $\gx$. It turns out that the corresponding models are indeed well-behaved\footnote{No subtleties occur in this case, since the model has a
non-singular vacuum. } for a 
range of non-zero values of this parameter as show below.  

To determine the physical realization and the spectrum of the theory, we have to minimize the potential 
\begin{eqnarray}
V &=& \frac {g_1^2}{10} (\gx -i\cM_Y)^2 + \frac{g_5^2}{2} \tr (-i\cM_t)^2.
\labl{SU5xU1}
\end{eqnarray}
This potential has 
absolute minimum at zero if
\begin{eqnarray}
|z|^2 &=& | k |^2 = 0, \quad -\frac{5}{2f^2}\leq \xi < 0,\nonumber\\
| h |^2 &=& \frac{1}{4f^2} +  
\frac{1}{10}\gx = v^2.
\labl{VEV}
\end{eqnarray}
This solution is supersymmetric and spontaneously breaks $U(1)$, whilst 
$SU(5)$ is manifestly preserved. As a result, the $U(1)$ gauge field 
$\tilde{A}$ become massive with a mass $\mathrm{m}^2_{\tA} = 
\mathrm{m}^2_\gr$ and the remaining vectors $V_\gm$ of $SU(5)$ together with Majorana fermions $\gl_{R}^i\,_{j}$ that are the gauginos of unbroken $SU(5)$ symmetry stay massless. The right-handed components of the gaugino $\tilde{\gl}_R$ combine with the left-handed chiral fermions $\varphi_L$ to 
become massive Dirac fermions with the same  mass as the gauge boson 
$\tilde{A}$: $\mathrm{m}^2_{\tA} = 
\mathrm{m}^2_\gr = \mathrm{m}^2_{\gPs} = 40 g^2_{1}v^2$. This establishes the presence of a massive vector 
supermultiplet. 

We end this part of the letter by remarking that one can also consider gauging  either 
the $U(1)$ $(g_5 = 0 )$ or $SU(5)$  $(g_1 = 0 )$ symmetry. In the first 
case when gauging only the $U(1)$ symmetry, the minimum potential is at the 
same point as in the $SU(5)\times U(1)$ gauging. Therefore the above 
discussion applies here and one gets the same spectrum with equal masses 
for the $U(1)$ gauge multiplet. On the other hand, if  only $SU(5)$ is gauged, the potential reaches its
 minimum at $z = k= 0$. Then no supersymmetry breaking or internal symmetry 
breaking occurs and all particles in the theory are massless.\\

We now turn our attention to another well known model with a 
phenomenologically interesting particle spectrum,
defined on the homogeneous coset space $E_6/SO(10) \times U(1)$ 
\cite{ysj1,YacSaoJv}.  The target manifold $E_6/SO(10) \times U(1)$ is 
parametrized by 16 complex fields $z^{\ga}$ in a chiral superfield $\gF_\ga = 
(z_\ga,\gps_{L\ga}, H_\ga)$ ($\ga = 1, ..., 16 $), 
transforming as a Weyl spinor under $SO(10)$. Their chiral fermion 
superpartners have the quantum numbers of one full generation of 
quarks and leptons, including a right-handed neutrino. To cancel the $U(1)$-anomaly 
the model is extended to a complete $\underline{27}$ of $E_6$. According to the 
branching rule:
$\underline{27}\, \rightarrow\, \underline{16}(1) + 
 \underline{10}(-2) + \underline{1}(4)$, where the numbers in parentheses 
denote the relative $U(1)$ 
weights. With this choice of matter content, the cancellation of 
chiral anomalies of the full $E_6$ isometry group is achieved \cite{SJ1} by 
introducing a 
superfield $\gPs_m = (N_m, \gc_{Lm})$ ($m = 1,\dots, 10$) which is equivalent to a
$\underline{10}$ of $SO(10)$ with $U(1)$ 
charge -2; and finally a singlet $\gL =(h, \gc_L)$ of $SO(10)$, with $U(1)$ 
charge +4. 

The anomaly-free supersymmetric $\gs$--model 
on $E_6/[SO(10)\times U(1)]$, is then defined by three chiral superfields $(\gF_\ga, \Psi_m, \gL)$ with \Kh\ 
potential given by
\begin{eqnarray}
\cK(\gF,\bgF; \gPs,\bgPs; \gL,\bgL) &=& K_\gs  + e^{-6f^2K_\gs}|h|^2\nonumber\\
& & 
+ g_{mn}\bN_m N_ne^{6f^2K_\gs},
\labl{7.1}
\end{eqnarray}
%
with $K_\gs = \bz .[Q^{-1}\ln (1 + Q)].z$, the $\gs$-model \Kh\  potential. We have introduced a constant $f$ with the dimension $m^{-1}$, determining 
the scale 
of symmetry breaking $E_6\rightarrow SO(10)\times U(1)$. The positive definite matrix $Q$ 
is defined as
\begin{eqnarray}
Q_\ga\,^\gb &=& \frac{f^2}{4}M_{\ga\gg}^{\gb\gd}\bz^{\gg}z_\gd,\nonumber\\
M_{\ga\gg}^{\gb\gd} &=& 3\gd_\ga^{+\gb}\gd_\gg^{+\gd} - 
\frac{1}{2}\gG_{mn\ga}^{+\,\,\,\,\,\,\gb}\gG_{mn\gg}^{+\,\,\,\,\,\,\gd}.
\end{eqnarray}
%
Here $\gG_{mn}^+  = \gG_{mn}\gd^+$ are the generators of the $SO(10)$ 
on positive chirality spinors of $SO(10)$ \cite{ysj1}, and $\gd^+$ is the 10-D positive 
 chirality projection operator. Furthermore $g_{mn}$ is the induced metric for the 10-vector 
representation defined by 
\begin{eqnarray}
g_{mn} = \frac{1}{16}\tr\Bigl(g_T(\gS_m C)^\dag g_T(\gS_n C)\Bigl),
\end{eqnarray}
with $g_T = (\Id_{16} + Q)^{-2}$.
The lagrangian constructed from the \Kh\ potential (\ref{7.1}) is invariant under 
a set of holomorphic Killing vectors generating a non-linear representation 
of $E_6$:
\begin{eqnarray}
\gd z_\ga &=& \frac{i}{2}\gth\sqrt{3}z_\ga 
- \frac{1}{4}\go_{mn}(\gG^+_{mn}\cdot z)_\ga + \frac{1}{2}\Bigl[
\frac{i}{f}\ge_\gb\gd^\gb_\ga\nonumber\\
&& 
- \frac{if}{4}\bge^\gb M^{\gg\gd}_{\ga\gb}z_\gg z_\gd\Bigl],\quad
\nonumber\\
[2mm]
\gd h &=& 2 i \Bigl(\sqrt{3}\gth - 3 f \bge\cdot z\Bigl)h,\nonumber\\
[2mm]
\gd N_n &=& - i\sqrt{3}\gth N_n - \go_{nm}N_m - i f\bge\cdot (\gG^+_{mn}\nonumber\\[2mm]
&& - 
3\gd^+_{mn})\cdot z N_m.
\labl{killE6}
\end{eqnarray} 
%
%
where $\gd^+_{mn} = \gd_{mn}\gd^+$, and $\gth$, $\go_{mn}$, $\ge_\ga$ and $\bge^\ga$  are the 
infinitesimal parameters of the $U(1)$, $SO(10)$ and broken 
$E_6$ generators respectively. The corresponding Killing potentials are 
\begin{eqnarray}
\cM_i &=& 
M_i\,E\\
[2mm]
& & - \frac{1}{8}e^{6f^2K_\gs}M_{i,\ga}^{~~\gb}\,g_{T\gg}^{~~\gd}(C\bgS_m)^{\ga\gg}
(\gS_n C)_{\gb\gd}\bN_m N_n\nonumber,
\end{eqnarray}
with $E$ and the $\gs$-model Killing potentials $M_i = (M_\gth, M^{(mn)}, \bM^\gb, M_\gb )$ given by 
\begin{eqnarray}
M_\gth &=& \frac{1}{f^2\sqrt{3}} - \frac{1}{2}\sqrt{3}\bz^\ga K_{\gs,\ga}\quad
\bM^\gb  = -\frac{1}{f}K_{\gs,}\,^\gb
,\nonumber\\[2mm] 
M^{mn} &=& - \frac{i}{2}\bz^\ga\gG^{+}_{mn\ga}\,^\gb K_{\gs,\gg},\quad M_\gb = -\frac{1}{f}K_{\gs,\gb},\\[2mm]
E &=& 1 - 6f^2e^{-6f^2K_\gs}|h|^2 + 6f^2e^{6f^2K_\gs}g_{mn}\bN_m N_n\nonumber.
\end{eqnarray}
%
%
 
Apart from the pure supersymmetric $\gs$--model determined by this 
\Kh\ potential (\ref{7.1}), we consider models in which (part of) the
isometries (\ref{killE6}) are gauged. As the $E_6$ is broken, the Higgs 
mechanism operates as follows: the Goldstone 
bosons $(\bz^\uga, z^\ga)$ are absorbed in 
the longitudinal component of the charged vector bosons, and if the full $E_6$ is 
gauged, we may choose the unitary gauge $\bz^\uga = z_\ga = 0$. To analyze the model 
in this gauge, we introduce the covariant derivatives for
the dynamical fields. The expressions for gauge-covariant derivatives of the 
complex scalar fields read
\begin{eqnarray}
D_\gm z_\ga &=& \der_\gm z_\ga - g\Bigl(\frac{i}{2}\sqrt{3}z_\ga A_\gm + 
\frac{1}{4}(\gG^+_{mn}z)_\ga A_{\gm(mn)}\nonumber\\
[2mm]&&
 + \frac{1}{2}(
\frac{i}{f}A_{\ga\gm}
-\frac{if}{4}\bA_\gm^\gb\,M^{\gg\gd}_{\ga\gb}\,z_\gg z_\gd)\Bigl),
\nonumber\\
[2mm]
D_\gm h &=& \der_\gm h - 2 i g\Bigl(\sqrt{3}A_\gm - 
3 f \bA^\ga_\gm z_\ga\Bigl)h,
\nonumber\\
[2mm]
D_\gm N_n &=& \der_\gm N_n +  i \sqrt{3} g A_\gm N_n + g A_{\gm(mn)} N_m\nonumber\\
[2mm]
& & + i f g \bA_\gm\cdot (\gG^+_{mn} - 
3\gd^+_{mn})\cdot z N_m.
\labl{E6gcovd}
\end{eqnarray}
%
%
Here we have introduced the notation $(A_{\gm\ga},\bA^\ga_\gm)$ for the 32 
charged gauge fields 
corresponding to the broken $E_6$ transformations; $A_{\gm(mn)}$ and 
$A_\gm$ are the gauge fields for the remaining $SO(10)$ and $U(1)$ 
transformations respectively. 

Next we discuss in some detail the gauging of the full non-linear $E_6$ and the resulting particle spectrum of the theory. We choose to study the model in the  unitary 
gauge in which all the Goldstone bosons vanish: $z^\ga = \bz_\ga = 0$. This implies 
that the broken Killing potentials $\bcM^\gb$ and $\cM_\gb$ vanish automatically, leaving us with $SO(10)$ and $U(1)$ Killing potentials $\cM_\gth$ and $\cM_{mn}$; and the $D$-term potential reads
\begin{equation}
V_D = \frac{g^2}{2}\Bigl(\cM^2_\gth + 
\frac{1}{2}\cM^2_{mn}\Bigl).
\labl{fullgE6}
\end{equation}
In the unitary gauge the potential becomes
\begin{eqnarray}
V_{\mathrm{unitary}} &=& 
\frac{g^2}{2}\Bigl(\frac{1}{f^2\sqrt{3}} - 2\sqrt{3}|h|^2 + 
\sqrt{3}\sum_m|N_m|^2\Bigl)^2\nonumber\\
[2mm]
& & + \frac{g^2}{2}\sum_{m,n}|\bN_m N_n - 
\bN_n N_m|^2.
\labl{fullgE61}
\end{eqnarray}
%
%
Minimization of the potential leads to the following set of supersymmetric 
minima characterized by the equation
\begin{eqnarray}
|\bN_m N_n - \bN_n N_m|^2 = 0,\quad |h|^2 = \frac{1}{6f^2} + 
\frac{1}{2}\sum_m|N_m|^2.
\labl{singularmin}
\end{eqnarray}
The value of the potential vanishes: $\langle V \rangle = 0$, hence it is the absolute minimum of the potential. From (\ref{singularmin}), it follows that $|h|\neq 0$ and the $U(1)$ gauge symmetry is always broken; a solution 
with $|N_m| = 0$ is possible, preserving $SO(10)$. However, this solution is not acceptable by itself, as it leads to the 
to the vanishing of the metric of the $\gs$--model fields 
\begin{eqnarray}
\langle G_\ga\,^\gb\rangle &=& \gd_{\ga}^{~\gb}\Bigl(\frac{1}{f^2} - 6 |h|^2 + 18 |N_m|^2 \Bigl) - 4
\bN_m N_n(\gG_{mn}^+)_\ga^\gb\nonumber\\
[2mm]
& =& 0.
\end{eqnarray} 
Hence the masses of the 32 $E_6$ gauge fields $A^\ga_\gm$ vanish, and the four-fermion term $R_{z^\ga\bz^\ugb h\bh}\,\bgps^\ga_R\,\gps^\ugb_L\,\bgc_L\,\gc_R$
diverges, just like in the $SO(10)/U(5)$--spinor model.
Clearly, in this domain the model no longer 
correctly describes the physics of the situation (i.e., the correct vacuum 
and the corresponding spectrum of small fluctuations). Therefore we add soft breaking
mass terms for the singlet $h$ and the vector $N_m$:
\begin{eqnarray}
V_{\mathrm{soft}} = \gm^2_1\,e^{-6K_\gs}|h|^2 + \gm^2_2\,g_{mn}\,\bN_m N_n\,e^{6f^2K_\gs},
\labl{softbreatems}
\end{eqnarray} 
to shift the minimum a way from the singular point. The full scalar potential with soft breaking term in the unitary gauge is then:
\begin{eqnarray}
V = V_{\mathrm{unitary}} + \gm^2_1 |h|^2 + \gm^2_2 |N_m|^2.
\labl{vfuul}
\end{eqnarray} 

We now determine the mass spectrum of the theory. As the complex scalar transforms only under $U(1)$, we choose the unitary gauge 
for the $U(1)$ symmetry, which allow us to write 
\begin{eqnarray}
h = \Bigl(v + \frac{1}{\sqrt{2}}\gr\Bigl)\,e^{\frac{1}{\sqrt{2v}}i\gk},
\labl{u1unitary}
\end{eqnarray} 
where $\gk$ is the longitudinal component of the massive gauge field 
$A_\mu$. Expanding the potential (\ref{vfuul}) to second order in $\gr$ and in the complex fluctuations $\tilde{N}_m$ around the minimum
\begin{eqnarray}
|N_m|^2 = 0,\quad |h|^2 = v^2 = \frac{1}{6f^2} - \frac{\gm^2_1}{12 g^2}\quad\gm^2_1 < 2\frac{g^2}{f^2}
\labl{softvacuum}
\end{eqnarray} 
one obtains the following bosonic mass spectrum: 
\begin{eqnarray}
{\mathrm{m}}^2_{\tA} &=& {\mathrm{m}}^2_\rho = 24g^2 v^2,\quad{\mathrm{m}}^2_{A^\ga} 
= \frac{\gm^2_1}{4f^2},\nonumber\\
[2mm]
{\mathrm{m}}^2_{\tilde{N}_m} &=& \frac{1}{f^2}\Bigl(\frac{1}{2}
\gm^2_1 + \gm^2_2\Bigl),\quad {\mathrm{m}}^2_{A_{nm}} = 0.
\end{eqnarray}
As expected the gauge bosons  $A_{\gm[mn]}$ of the non-broken 
$SO(10)$ symmetry remain massless. Analyzing the kinetic and mass terms of the fermions one realizes that two massive Dirac fermions can be formed by combining the fermions of the chiral multiplets with two 
gauginos:
\begin{eqnarray}
\gPs_\ga = \frac{1}{\sqrt{2}}\gl_{R\ga} - i\sqrt{2}\,\frac{\gm_1}{2g}\,\gps_{L\ga}, 
\qquad 
\gO = \gl_R - i\gc_L.
\labl{Omega}
\end{eqnarray}
The masses of these spinors read:
$\mathrm{m}^2_{\gPs} = \frac{\gm^2_1}{2f^2}$ and $\mathrm{m}^2_{\gO} = 
24g^2 v^2$.
The $\underline{16}$ of the left-handed gaugino's $\gl_{L\ga}$ and 
quasi-Goldstone fermions $\gc_{Ln}$ remain massless, together with the Majorana fermions $\gl^{mn}$ that are 
gauginos of the unbroken $SO(10)$ symmetry. Therefore, in this model the gaugino 
components $\gl_{L\ga}$ are now to be identified with a family of quarks and leptons, 
rather than the quasi Goldstone fermions themselves. (We have observed a similar thing to happen also in the $SO(10)/U(5)$--spinor model discussed earlier.) 

The gauging of the $SO(10)\times U(1)$ symmetry instead of the full $E_6$ 
gives analogous, but not quite identical, results. Also in this case one 
finds 
the potential (\ref{fullgE6}), but in general with different values $g_1$ and 
$g_{10}$ for the coupling constants of $SO(10)$ and $U(1)$. Except for 
special values of the parameters, it has a minimum for the $SO(10)$ invariant 
solution, with $\langle z_\ga\rangle = 0$; and again the metric becomes singular. One way 
to shift the minimum away from this point is by introducing soft breaking 
terms (\ref{softbreatems}). Another option is to add an extra 
Fayet-Iliopoulos term as the gauge group possesses an explicit $U(1)$ 
factor.  In the first case, one obtains similar mass spectra as in the model with 
fully gauged $E_6$, but with $g = g_1$ and only one massive Dirac fermion 
$\gO$ defined in (\ref{Omega}). The gauginos $\gl^{mn}$ that are left over remain unpaired, and hence 
massless. Furthermore, the chiral fermions  $\gps^\ga_L$ and $\gc^n_L$  
remain massless. In the second case, for special values of the coupling constants 
$g_1$ and $g_{10}$, or 
the Fayet-Iliopoulos parameter $\gx$, one can get different results. 
Since the 
$SO(10)$ and $U(1)$ coupling constants are independent, one may choose to gauge only  
$SO(10)$ ($g_1 =  0$). In that case both supersymmetry and  internal symmetry 
are preserved, and the particle spectrum of a model contains of a massless $SO(10)$ 
gauge boson, just like in the usual supersymmetric $SO(10)$ grand unified models.

Solutions with $|N_m|\neq 0$ breaking $SO(10)$ are allowed, and expected in the 
next stage of the symmetry breaking. For example, $SO(10)$ broken solution can be 
chosen as 
\begin{eqnarray}
f\bN_m &=& \left(
\begin{array}{cccccccccc}
0&0&0&0&0&0&0&0&0&v_{10}
\labl{n10}
\end{array}
\right),\nonumber\\
[2mm]
|h|^2 &=& |v_h|^2 = \frac{1}{6f^2} + \frac{v^2_{10}}{2f^2}.
\end{eqnarray}
Since the complex scalar $N_m$ gets a vacuum expectation value; this breaks 
the internal linear $SO(10)$ symmetry, leaving only $SO(9)$. As  
supersymmetry is preserved, one expects the spectrum of 
physical states to fall into supersymmetric multiplets with vanishing mass 
supertrace. Indeed the general mass sum rule\footnote{As we have gauge the full $E_6$ the standard linear Fayet--Iliopoulos 
term is of course absent.} \cite{Grisaru,Gates} leads to
\begin{eqnarray}
{\rm STr}\,\rm{m}^2 &=&  2 g^2\,G^{\uI I}\,\cM_i\,\cM_{i,\uI I} = 0.
\labl{sumruleE6}
\end{eqnarray}\\
\nit
In this paper, I have discussed in some detail the analysis of of particle spectrum 
of supersymmetric $\gs$-models on homogeneous  coset-spaces $E_6/[SO(10)\times U(1)]$ 
and $SO(10)/U(5)$. I have analyzed the possible vacuum configurations of 
these models and investigate the existence of the zeros 
of the potential, for which the models are anomaly-free, with positive definite 
kinetic energy.  The consequences of these physical requirements have been 
analyzed. We found that there exist supersymmetric minima for both these models when the full isometry groups $E_6$ and $SO(10)$ are gauged. The analysis is straightforward 
as one can employ the unitary gauge to put the Goldstone bosons to zero. In some cases, we find that the \Kh\ metrc is singular: the kinetic energy of the would-be Goldstone modes and their fermionic partners vanishes in the vacuum. We showed by addition of soft supersymmetry-breaking mass parameters, that 
the minimum can be shifted away from the singular point. The particle spectrum in the presence of soft supersymmetry-breaking mass parameters is computed.

Continuing our line of investigation of the particle spectrum of supersymmetric 
$\gs$-models on $E_6/[SO(10)\times U(1)]$, and $SO(10)\times U(1)$, we have also studied the 
possibility of gauging (part of)  the linear subgroups, i.e., $SO(10)\times U(1)$ 
and $U(5)$. In each of these models, we found that the  properties of the 
model investigated depend to a certain extent on the value of 
parameters (gauge couplings, Fayet-Iliopoulos term.)  We have obtained all supersymmetric minima, of which some are 
physically problematic as the kinetic terms of the Goldstone multiplets either 
vanish or have negative values.\\

\nit
I thank J.W.\ van Holten and J. Babington for very useful suggestions and for reading the manuscript.  

\end{document}